\documentclass[onecolumn,showpacs,fleqn,nobibnotes]{revtex4}

\usepackage{amsmath}
\usepackage{graphicx}
\usepackage{float}
\usepackage{subfigure}

\def\lsim{\raise0.3ex\hbox{$<$\kern-0.75em\raise-1.1ex\hbox{$\sim$}}}
\def\gsim{\raise0.3ex\hbox{$>$\kern-0.75em\raise-1.1ex\hbox{$\sim$}}}

\newcommand{\beq}{\begin{equation}}
\newcommand{\eeq}{\end{equation}}

\begin{document}

\title{Inelastic quarkonium photoproduction in coherent hadron - hadron  interactions at LHC energies}
\author{V.P. Gon\c{c}alves$^{a}$ and M.M. Machado$^{b}$}

\affiliation{$^{a}$ Instituto de F\'{\i}sica e Matem\'atica, Universidade Federal de
Pelotas\\
Caixa Postal 354, CEP 96010-900, Pelotas, RS, Brazil.}

\affiliation{$^{b}$ Instituto Federal de Educa\c{c}\~ao, Ci\^encia e Tecnologia, IF - Farroupilha,
Campus S\~ao Borja \\ Rua Otaviano Castilho Mendes, 355, CEP 97670-000, S\~ao Borja, RS, Brazil.}

\begin{abstract}
In this paper we study the inelastic quarkonium photoproduction in coherent $pp/pPb/PbPb$ interactions. Considering the ultra relativistic hadrons as a source of photons, we estimate the  total    $ h_1 + h_2 \rightarrow h \otimes V + X$ ($V = J/\Psi$ and $\Upsilon$) cross sections and rapidity distributions at LHC energies. Our results  demonstrate that the experimental analysis of this process can be used to understand the underlying mechanism governing heavy quarkonium production.
\end{abstract}
\maketitle

\section{Introduction}
\label{intro}

In the last years, the analysis of coherent hadron-hadron collisions becomes  an alternative way to study the theory of strong interactions - the Quantum Chromodynamics (QCD) - 
in the regime of high energies (For  a review see Ref. \cite{upc}).
The basic idea in coherent  hadronic collisions is that the total cross section for a given process can be factorized in terms of the equivalent flux of photons { in} the hadron projectile and the photon-photon or photon-hadron production cross section. The main advantage of using colliding hadrons and nuclear beams for studying photon induced interactions is the high equivalent photon energies and luminosities, that can be achieved at existing and future accelerators. In particular, the photon-hadron interactions can be divided into exclusive and inclusive reactions. In the first case, a certain particle is produced, while the target remains in the ground state (or is only internally excited). On the other hand, in inclusive interactions the particle produced is accompanied by one or more particles 
from the breakup of the target. The typical examples of these processes are the exclusive vector meson production, described by the process 
$\gamma h \rightarrow V h$ ($V =  J/\Psi, \Upsilon$), and the inclusive heavy quark production [$\gamma p \rightarrow X Y$ ($X = c\overline{c}, b\overline{b}$)], 
respectively (For related discussions see, e.g. Refs. \cite{vicz0,vicmairon,vicwerner}). 
Recent experimental results from CDF  \cite{cdf} at Tevatron, STAR \cite{star} and PHENIX \cite{phenix} at RHIC and ALICE \cite{alice} and LHCb \cite{lhcb} at LHC have demonstrated that the study of coherent interactions in these colliders is feasible and that the data can be used to constrain the description of the hadronic structure at high energies.  It motivates the analysis of the production of other final states.

\begin{figure*}[t]
\includegraphics[scale=0.39]{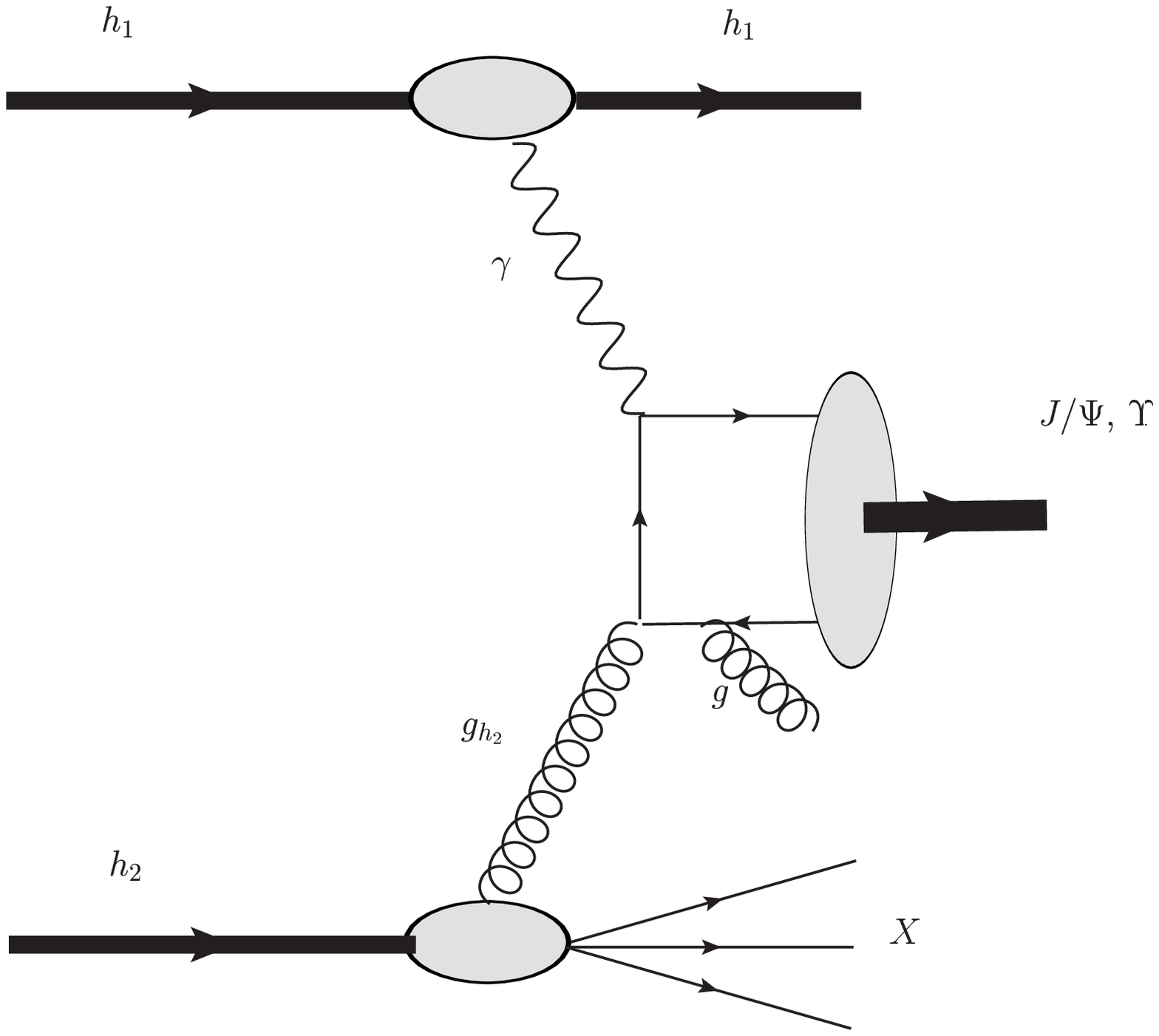} 
\hspace{0.6cm}
\includegraphics[scale=0.39]{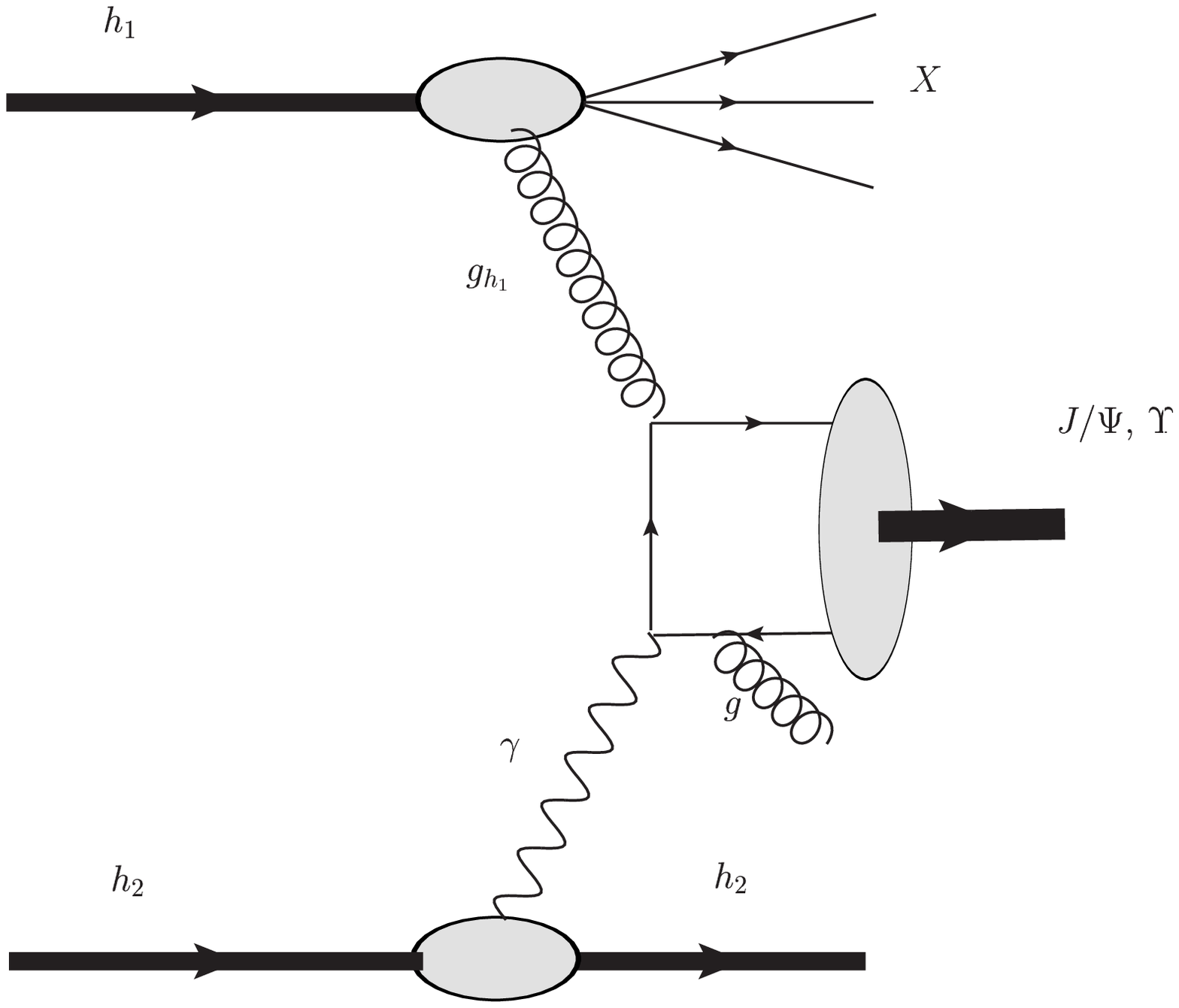}
 \caption{The mechanism for the inelastic quarkonium photoproduction  in coherent hadron - hadron interactions.  }
\label{fig:0}
\vspace{1cm}
\end{figure*}

One of the most interesting channels for high energy photon collisions at the LHC is the production of vector mesons such as heavy quarkonia. The exclusive vector meson production in coherent hadron - hadron interactions has been a subject of intense analysis in the last years \cite{gluon,klein_prc,strikman,vicmag_mesons,vicmag_mesons2,Schafer}, with a renewed motivation \cite{vicmag_update,Schafer2,gluon2,Lappi,griep,Guzey,Martin} associated to the recent  experimental data from ALICE and LHCb collaborations. However, it is well known that heavy quarkonia  can also be produced in inclusive photoproduction processes, in which the hadron target dissociates, and that the corresponding cross section grows with the energy. The contribution of this process in coherent interactions still is an open question.     
In this paper we study, for the first time, the inelastic  quarkonium production in coherent $pp/pPb/PbPb$ interactions. 
The main motivation for our analysis is to verify if it is possible to do an experimental study of this channel  at LHC in order to elucidated the poorly understood inclusive quarkonium mechanism. Our goal is to obtain an estimate of the total cross sections for $J/\Psi$ and $\Upsilon$ production and the corresponding rapidity distributions for the inelastic production. This process can be considered a background of the exclusive production if we take into account the large event pileup expected at LHC, which difficult the observation of the rapidity gaps in the final states. Consequently, the separation between these two processes should be possible at LHC only if the hadrons in the final states were tagged. Such possibility is currently under study (See e.g. \cite{afp}).

A schematic view of the mechanism which we consider in this paper is presented in Fig. \ref{fig:0}. Distinctly from the exclusive production, which is characterized by two rapidity gaps in the final state, in the inclusive case we  have only one rapidity gap, associated to the photon exchange.   Moreover, the photoproduction cross section   in the inclusive is linearly proportional to the gluon distribution ($xg$), while in the exclusive case it is proportional to  $xg$ squared
The cross section for the  inelastic quarkonium photoproduction  in a coherent hadron-hadron collision  is given by,
\begin{equation}
   \sigma(h_1+h_2 \rightarrow h \otimes V + X) =  \int d \omega \frac{dN}{d\omega}|_{h_1}
   \, \sigma_{\gamma h_2 \rightarrow V X}\left(W_{\gamma h_2}  \right) + \int d \omega \frac{dN}{d\omega}|_{h_2}
   \, \sigma_{\gamma h_1 \rightarrow V X}\left(W_{\gamma h_1}  \right)\,  \; , 
\label{eq:sigma_pp}
\end{equation}
where $\otimes$ represents the presence of one rapidity gap in the final state, $\omega$ is the photon energy in the center-of-mass frame (c.m.s.), $\frac{dN}{d\omega}|_{h_i}$ is the equivalent photon flux for the hadron $h_i$, $W_{\gamma h}$ is the c.m.s. photon-hadron energy given by $W_{\gamma h}^2=2\,\omega\sqrt{s}$, where
$\sqrt{s}$ is  the c.m.s energy of the
hadron-hadron system. Considering the requirement that  photoproduction
is not accompanied by hadronic interaction (ultra-peripheral
collision) an analytic approximation for the equivalent photon flux of a nuclei can be calculated, which is given by \cite{upc}
\begin{eqnarray}
\frac{dN}{d\omega}|_{A}= \frac{2\,Z^2\alpha_{em}}{\pi\,\omega}\, \left[\bar{\eta}\,K_0\,(\bar{\eta})\, K_1\,(\bar{\eta})- \frac{\bar{\eta}^2}{2}\,{\cal{U}}(\bar{\eta}) \right]\,
\label{fluxint}
\end{eqnarray}
where   $K_0(\eta)$ and  $K_1(\eta)$ are the
modified Bessel functions, $\bar{\eta}=\omega\,(R_{h_1}+R_{h_2})/\gamma_L$ and  ${\cal{U}}(\bar{\eta}) = K_1^2\,(\bar{\eta})-  K_0^2\,(\bar{\eta})$.
 On the other hand, for   proton-proton interactions, we assume that the  photon spectrum is given by  \cite{Dress},
\begin{eqnarray}
\frac{dN}{d\omega}|_{p} =  \frac{\alpha_{\mathrm{em}}}{2 \pi\, \omega} \left[ 1 + \left(1 -
\frac{2\,\omega}{\sqrt{S_{NN}}}\right)^2 \right] 
\left( \ln{\Omega} - \frac{11}{6} + \frac{3}{\Omega}  - \frac{3}{2 \,\Omega^2} + \frac{1}{3 \,\Omega^3} \right) \,,
\label{eq:photon_spectrum}
\end{eqnarray}
with the notation $\Omega = 1 + [\,(0.71 \,\mathrm{GeV}^2)/Q_{\mathrm{min}}^2\,]$ and $Q_{\mathrm{min}}^2= \omega^2/[\,\gamma_L^2 \,(1-2\,\omega /\sqrt{S_{NN}})\,] \approx (\omega/
\gamma_L)^2$, where $\gamma_L$ is the Lorentz boost  of a single beam. This expression  is derived considering the Weizs\"{a}cker-Williams method of virtual photons and using an elastic proton form factor (For more details see Refs. \cite{Dress,Kniehl}). 
Equation (\ref{eq:sigma_pp}) takes into account the fact that the incoming hadrons can act as both target and photon emitter.  

The main input in our calculations is the inelastic quarkonium photoproduction  cross section, $\sigma_{\gamma + h \rightarrow V + X}$, which we estimate in the Section \ref{quarkonium} considering the Color Singlet Model \cite{berger}. A comparison with  the $ep$ HERA data is also presented. Moreover, in Section \ref{results} we present our predictions for the rapidity distributions and total cross sections for $J/\Psi$ and $\Upsilon$ production at LHC energies and in Section \ref{conc} we summarize our main conclusions.

\begin{figure}
\includegraphics[scale=0.18]{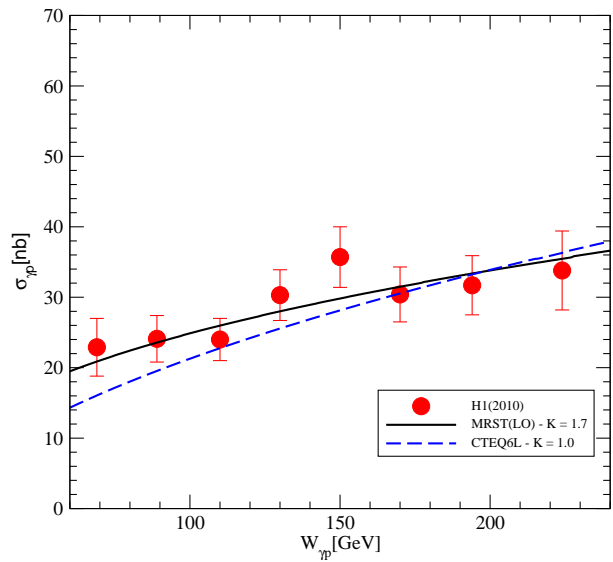}
\caption{ Energy dependence of the   inelastic $J/\Psi$ cross section considering two different parameterizations for the gluon distribution. Data from H1 Collaboration \cite{h1} for $0.3 < z < 0.9$ and $p_{T} > 1$ GeV.}
\label{fig1}
\end{figure}

\section{The  Inelastic Quarkonium Photoproduction }
\label{quarkonium}

The study of the production of heavy quarkonium states provides a unique laboratory in which to explore the interplay between perturbative and nonperturbative effects in QCD (For a recent review see, e.g., Ref. \cite{review_nrqcd}).
In the last decades, a number of theoretical approaches have been proposed for the calculation of these states, as for instance,  the Non Relativistic QCD (NRQCD)  approach, the fragmentation approach, the color singlet model (CSM), the color evaporation model and the $k_T$-factorization approach.
In the NRQCD formalism \cite{nrqcd} the cross section for the production of a heavy quarkonium state $V$ factorizes as  $\sigma (ab \rightarrow V+X)=\sum_n \sigma(ab \rightarrow Q\bar{Q}[n] + X) \langle {\cal{O}}^V[n]\rangle$, where the coefficients $\sigma(ab \rightarrow Q\bar{Q}[n] + X)$ are perturbatively calculated short distance cross sections for the production of the heavy quark pair $Q\bar{Q}$ in an intermediate Fock state $n$, which does not have to be color neutral.  The $\langle {\cal{O}}^V[n]\rangle$
are nonperturbative long distance matrix elements (LDME), which describe the transition of the intermediate $Q\bar{Q}$ in the physical state $V$ via soft gluon radiation. Currently, these elements have to be extracted in a global fit to quarkonium data as performed, for instance, in Ref. \cite{bute}. 
In the Color Singlet Model \cite{berger}, only those states with the same quantum numbers as the resulting quarkonium contribute to the formation of a bound $Q\bar{Q}$ state. This is achieved by radiating a hard gluon in a perturbative process. In contrast, in NRQCD, also color octet $Q\bar{Q}$ states contribute to the quarkonium production cross section via soft gluon radiation. The Color Singlet Model can be obtained from NRQCD factorization by retaining, for a given process, only the contribution that is associated with the color-singlet LDME of the lowest non-trivial order in $v$, which is the typical velocity of the heavy quark or antiquark in the quarkonium rest frame. It is important to emphasize that the underlying mechanics governing heavy quarkonium production is still subject of intense debate \cite{review_nrqcd}.

\begin{figure}
\includegraphics[scale=0.3]{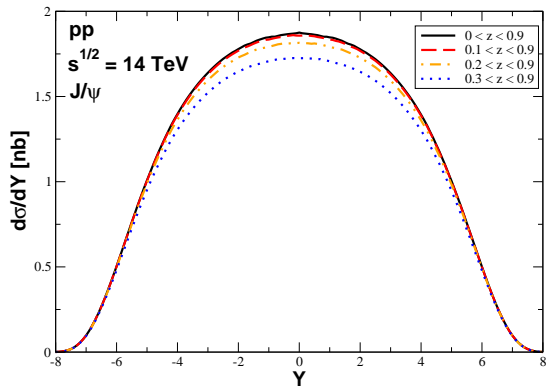}
\caption{(Color online). Dependence on $z_{min}$ of the rapidity distribution for the inelastic $J/\Psi$ photoproduction  in $pp$ collisions at $\sqrt{s} = 14$ TeV. }
\label{fig2}
\end{figure}

In the specific case of the  inelastic $V$ photoproduction, the description of the experimental data for this process is a challenge to the distinct approaches, as verified in Ref. \cite{h1} and discussed in detail in Refs. \cite{review_nrqcd,bk}.
In this paper we consider the CSM approach, which predicts that 
 the total cross section can be expressed at leading order  as follows (See e.g. \cite{ko})
\begin{eqnarray}
\sigma (\gamma + h \rightarrow V + X) = \int dz dp_{T}^2 \frac{xg_h(x,Q^2)}{z(1-z)} \frac{d\sigma}{dt}(\gamma + g \rightarrow V + g ) \label{sigmagamp}
\end{eqnarray}
where $z \equiv (p_V.p)/(p_{\gamma}.p)$, with $p_V$, $p$ and $p_{\gamma}$ being the four momentum of the quarkonium, hadron and photon, respectively. In the hadron rest frame, $z$ can be interpreted as the fraction of the photon energy carried away by the quarkonium. Moreover, $p_{T}$ is the magnitude of the quarkonium three-momentum normal to the beam axis.
The partonic differential cross section $d\sigma/dt$ is given by \cite{ko}
\begin{eqnarray}
\frac{d\sigma}{dt}(\gamma + g \rightarrow V + g) = \frac{64 \pi^2}{3}\frac{e_Q^4 \alpha^2\alpha_s m_Q}{s^2}
\left(\frac{s^2s_1^2+t^2t_1^2+u^2u_1^2}{s_1^2t_1^2u_1^2}\right)\langle O^{V}(^3S_1^{[1]})\rangle \label{dsdt}
\end{eqnarray}
where $e_Q$ and $m_Q$ are, respectively, the charge and mass of heavy quark constituent of the quarkonium. The Mandelstam variables can be expressed in terms of $z$ and $p_{T}$ as follows:
\begin{eqnarray}
s & = & \frac{p_{T}^2+(2m_Q)^2(1-z)}{z(1-z)}\,\,, \nonumber\\
t & = & - \frac{p_{T}^2+(2m_Q)^2(1-z)}{z}\,\,, \nonumber\\
u & = & - \frac{p_{T}^2}{1-z} \,\,.
\end{eqnarray}
Moreover, $s_1 = {s} - 4 m_Q^2$, $t_1 = {t} - 4 m_Q^2$,  $u_1 = u - 4 m_Q^2$ and the Bjorken variable $x$ can be expressed by
\begin{eqnarray}
x = \frac{p_{T}^2+(2m_Q)^2(1-z)}{W_{\gamma h}^2z(1-z)} \,\,,
\end{eqnarray}
	where $W_{\gamma h}$ is the photon - hadron center-of-mass energy.
The long distance matrix elements $\langle O^{V}(^3S_1^{[1]})\rangle$ can be determined from quarkonium electromagnetic decay rates. In our calculations 	
we use the values as given in Refs. \cite{ko} and \cite{wang} for the $J/\Psi$ and $\Upsilon$ production, respectively. Moreover, in what follows we  consider different parametrizations for the parton distributions. In particular, we use the MRSTLO \cite{mrst} and CTEQ6L \cite{cteq} parton distributions for the proton. In the nuclear case, we take into account the nuclear shadowing effects as given by the EPS09 parametrization \cite{eps09}, which is  based on a global fit of the current nuclear data.

\begin{figure}
\includegraphics[scale=0.37]{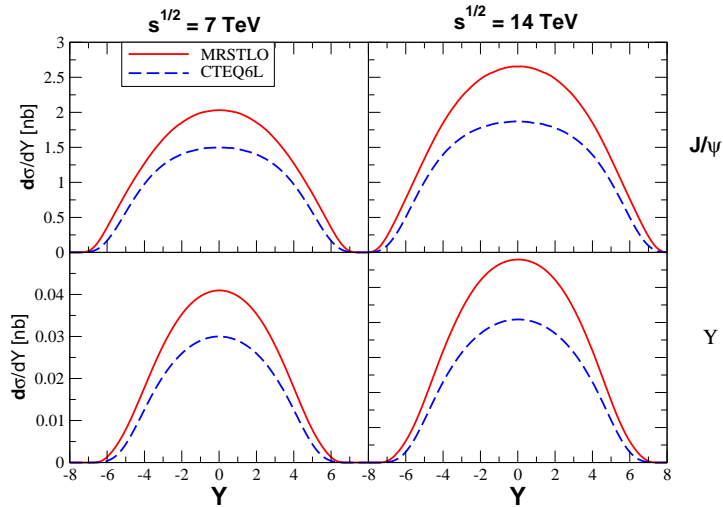}
\caption{Rapidity distribution for the $J/\Psi$ and $\Upsilon$ production  in coherent $pp$ collisions at $\sqrt{s} = 7$ TeV (left panels) and 14 TeV (right panels) considering two different parametrizations for the gluon distribution. }
\label{fig3}
\end{figure}

\begin{table}
\begin{center}
\begin{tabular} {||c|c|c||}
\hline
\hline
{\bf $J/\Psi$} & {\bf MRSTLO} & {\bf CTEQ6L}   \\
\hline
\hline
$\sqrt{s}=7$ TeV & 17.93 nb ($1793 \times 10^6$)  &  13.18 nb ($1318 \times 10^6$)   \\
\hline
$\sqrt{s}=14$ TeV & 25.66 nb ($2566 \times 10^6$)   &  18.40 nb ($1840 \times \,10^6$)     \\
\hline
\hline
{\bf $\Upsilon$} & {\bf MRSTLO} & {\bf CTEQ6L} \\
\hline
\hline
$\sqrt{s}=7$ TeV & 0.30 nb ($30 \times 10^6$)  & 0.21 nb ($21 \times 10^6$)    \\
\hline
$\sqrt{s}=14$ TeV & 0.47 nb ($47 \times 10^6$)  & 0.33 nb  ($33 \times 10^6$)  \\
\hline
\hline
\end{tabular}
\end{center}
\caption{The total cross section (event rates)  for the  inelastic quarkonium photoproduction in coherent $pp$ collisions at  LHC energies.}
\label{tab1}
\end{table}

A comment is in order. In our calculations we will consider the Color Singlet Model at leading order (LO) as given above. As demonstrated in Refs. \cite{kramer,kramer_plb} (See also \cite{maltoni}) the calculation of photoproduction cross section to next-to-leading order (NLO) proved that these corrections are large, increasing towards large transverse momentum of the $J/\Psi$ meson. The corresponding NLO predictions describes the shape  of the $z$ and $p_{T}$ differential $J/\Psi$ cross sections but the normalizations are a factor three below of data \cite{h1}, with large uncertainties  associated to the choice of the charm quark mass and the factorization and renormalization scales, which indicates that beyond NLO corrections should be included or that contributions of color octet states may be sizeable. In contrast, the total cross section can be described at LO by adjusting  the charm quark mass and factorization and renormalization scales and introducing a multiplicative $K$ - factor, which takes in account higher-order corrections. It is demonstrated in the Fig. \ref{fig1} where we compare our predictions, obtained using  $Q^2 = p_{T}^2$, $m_c = 1.5$ GeV and two different parametrizations for the gluon distribution, with the experimental data from H1 Collaboration \cite{h1}. We have that, considering distinct values for the $K$ - factor, both predictions reasonably describe the data. As the main input for the calculations of the inelastic $J/\Psi$ photoproduction in coherent $pp$ interactions is the energy dependence of the  $\gamma + p \rightarrow J/\Psi + X$ cross section, we believe that the use of Color Singlet Model at leading order, with parameters constrained by the HERA data, can be considered a reasonable first approximation for the total cross section and rapidity distribution. Certainly this subject deserves more detailed studies in the future. 
 In what follows we will assume that these same values of $Q^2$ and $K$-factor are valid for $J/\Psi$ production in $pPb$ and $PbPb$ collisions and also for $\Upsilon$ production with $m_b = 4.5$ GeV.


\begin{figure*}[t]
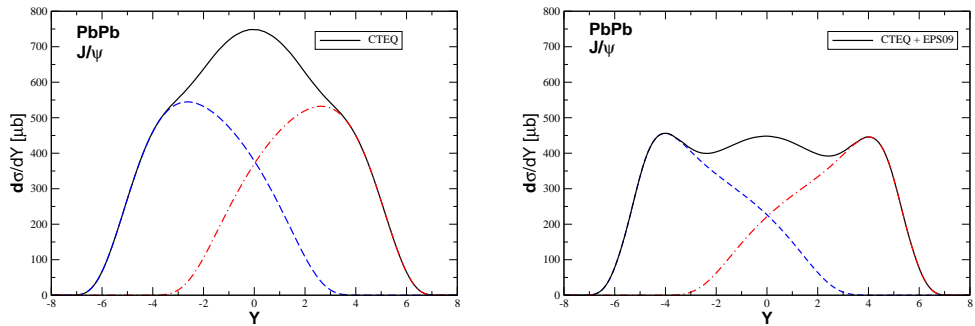

\includegraphics[scale=0.25]{PbPb_jpsi3.eps} 
\hspace{0.6cm}
\includegraphics[scale=0.25]{PbPb_jpsi2.eps}
 \caption{(Color online). Rapidity distribution for the inelastic $J/\Psi$ photoproduction in coherent $PbPb$ collisions at $\sqrt{s} = $ 5.5 TeV disregarding (left panel) and taken into account (right panel) the nuclear shadowing corrections. See text.}
\label{fig4}
\vspace{1cm}
\end{figure*}

\section{Results}
\label{results}

Lets calculate the rapidity distribution and total cross section for the inelastic quarkonium photoproduction in  coherent $pp$, $pPb$ and $PbPb$ collisions.
The distribution on rapidity $Y$ of the quarkonium in the final state can be directly computed from Eq. (\ref{eq:sigma_pp}), by using its  relation with the photon energy $\omega$, i.e. $Y\propto \ln \, ( \omega/M_{V})$.  Explicitly, the rapidity distribution is written down as,
\begin{eqnarray}
\frac{d\sigma \,\left[h_1 + h_2 \rightarrow   h \otimes V + X \right]}{dY} = \left[\omega \frac{dN}{d\omega}|_{h_1}\,\sigma_{\gamma h_2 \rightarrow V\, X}\left(\omega \right)\right]_{\omega_L} + \left[\omega \frac{dN}{d\omega}|_{h_2}\,\sigma_{\gamma h_1 \rightarrow V\, X}\left(\omega \right)\right]_{\omega_R}\,
\label{dsigdy}
\end{eqnarray}
where $\otimes$ represents the presence of a rapidity gap in the final state and $\omega_L \, (\propto e^{-Y})$ and $\omega_R \, (\propto e^{Y})$ denote photons from the $h_1$ and $h_2$ hadrons, respectively.  As the photon fluxes, Eqs. (\ref{fluxint}) and (\ref{eq:photon_spectrum}), have support at small values of $\omega$, decreasing exponentially at large $\omega$, the first term on the right-hand side of the Eq. (\ref{dsigdy}) peaks at positive rapidities while the second term peaks at negative rapidities. Consequently, given the photon flux, the study of the rapidity distribution can be used to constrain  the photoproduction cross section for a given energy. Moreover, in contrast to the total rapidity distributions for $pp$ and $PbPb$ collisions, which will be symmetric about midrapidity ($Y=0$), $d\sigma/dY$ will be asymmetric in $pPb$ collisions due to the differences between the fluxes and process cross sections.

Following Ref. \cite{h1} we will  integrate  the fraction of the photon energy carried away by the quarkonium in the  range  $0.3  \lesssim z \lesssim 0.9$. 
Our predictions do not include resolved photoproduction, which contributes appreciably only at $z \lesssim 0.3$ and diffractive production, which is confined to the quasielastic domain at $z \approx 1$ and $p_{T} \approx 0$. These contributions are in general excluded from experimental measurements (See Ref. \cite{h1} and Fig. \ref{fig1}) in order to make a meaningful comparison. In order to estimate the dependence of our results on the inferior limit of integration $z_{min}$, in Fig. \ref{fig2} we present our predictions for the rapidity distribution for $J/\Psi$ production in coherent $pp$ collisions at $\sqrt{s} = 14$ TeV obtained using the CTEQ6L parametrization and different values of $z_{min}$. We have that our predictions increase by $\approx$ 23 \% at midrapidity  if we assume $z_{min} = 0$. Similar behaviour is observed for other energies and for $\Upsilon$ production.

In Fig. \ref{fig3}  we present our predictions for the inelastic $J/\Psi$ and $\Upsilon$ photoproduction in coherent $pp$ collisions at LHC energies. As expected, the total rapidity distributions are symmetric about the midrapidity. We calculate $d\sigma/dY$ considering different parametrizations for the gluon distribution in the proton. 
It is important to emphasize that the rapidity distribution at LHC probes a large interval of photon-proton center of mass energy since $W^2_{\gamma h}\simeq M_{V}\,\sqrt{s}\,\exp({\pm Y})$, which corresponds to very small $x\simeq M_{V}\,e^{-|Y|}/\sqrt{s}$.  The MRSTLO and CTEQ6L predictions differ by $\approx$ 30 \%  at $Y=0$, with the CTEQ6L one being a lower bound.  In Table \ref{tab1} we present our estimates for the total cross sections and production rates
assuming  the  design luminosity ${\cal L}^{\mathrm{pp}}_{\mathrm{LHC}} = 10^7$ mb$^{-1}$s$^{-1}$ and a run time of $10^7$ seconds. We predict large values for the events rate and cross sections of the order of units of nb, in contrast with values of the order of $100$ $\mu$b for the inclusive quarkonium hadroproduction \cite{review_nrqcd}.  Despite their much smaller cross sections, the clean topology of  coherent processes  implies a larger signal to background ratio. Therefore, the experimental detection is in principle feasible.
However, as already emphasized in the Section \ref{intro} the signal is expected to be reduced due to the event pileup and an alternative  to  measure coherent events at the LHC is by tagging the intact hadron in the final state.

\begin{table}
\begin{center}
\begin{tabular} {||c|c|c||}
\hline
\hline
 & {\bf CTEQ} & {\bf CTEQ + EPS09}   \\
\hline
\hline
{\bf $J/\Psi$} & 6122.0 $\mu$b ($2571 \times 10^3$)    &  4647.6 $\mu$b ($1951 \times 10^3$)     \\
\hline
{\bf $\Upsilon$} & 71.9 $\mu$b ($30 \times 10^3$)  &  60.6 $\mu$b ($25 \times 10^3$)    \\
\hline
\hline
\end{tabular}
\end{center}
\caption{The total cross section (event rates)  for the  inelastic quarkonium photoproduction in coherent $PbPb$ collisions at  $\sqrt{s} = 5.5$ TeV.}
\label{tab2}
\end{table}

In what follows we will consider coherent $PbPb$ and $pPb$ collisions. We will  assume that the gluon distribution in the proton,  $xg_p$,  is given by the CTEQ6L parametrizations and that the nuclear gluon distribution is given by $xg_A = A.R_g.xg_p$, where $R_g$ takes into account nuclear shadowing effects as given by the EPS09 parametrization \cite{eps09}.
In Fig. \ref{fig4}  we present our predictions for the inelastic $J/\Psi$ photoproduction in coherent $PbPb$ collisions at $\sqrt{s} = 5.5$ TeV. In the left panel we present the predictions obtained disregarding the shadowing corrections ($R_g = 1$), while in right panel these corrections are taken into account. The two contributions of Eq. (\ref{dsigdy}) are presented separately as well as the sum (solid line). As in the $pp$ case, the total rapidity distributions are symmetric about the midrapidity. 
However, we obtain larger values for the rapidity distribution due to the enhancement of the photon flux for a nucleus, which is proportional to $Z^2$. As for the LHC energies  the typical  values of  $x$ are  $\simeq M_{V}\,e^{-|Y|}/\sqrt{s}$, we are probing the shadowing corrections ($R_g < 1$) in the nuclear gluon distributions. As we can see in Fig. \ref{fig4}, the behaviour of the rapidity distribution is strongly modified by these corrections. In Table \ref{tab2} we present our estimates for the total cross sections and event rates assuming  the  design luminosity ${\cal L}^{\mathrm{PbPb}}_{\mathrm{LHC}} = 0.42$ mb$^{-1}$s$^{-1}$.

\begin{figure*}[t]
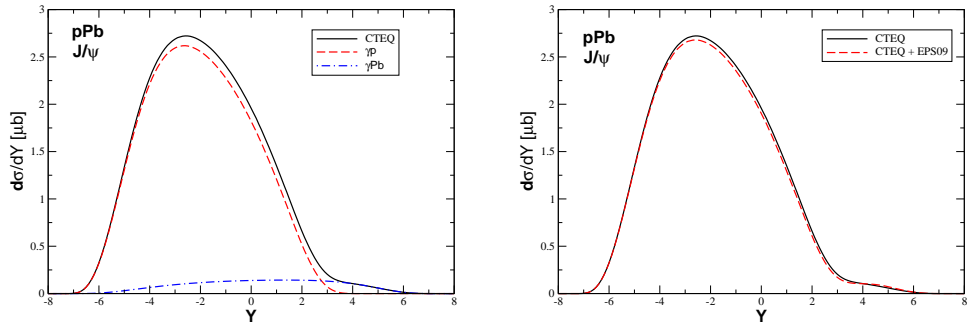

\includegraphics[scale=0.25]{pPb_jpsi.eps} 
\hspace{0.6cm}
\includegraphics[scale=0.25]{pPb_jpsi2.eps}
 \caption{(Color online). Rapidity distribution for the inelastic $J/\Psi$ photoproduction in coherent $pPb$ collisions at $\sqrt{s} = $ 5.5 TeV. The $\gamma p$ and $\gamma Pb$ contributions are explicitly presented in the left panel. In the right panel we present the predictions obtained disregarding and taken into account the nuclear shadowing corrections.}
\label{fig6}
\vspace{1cm}
\end{figure*}

\begin{figure*}[t]
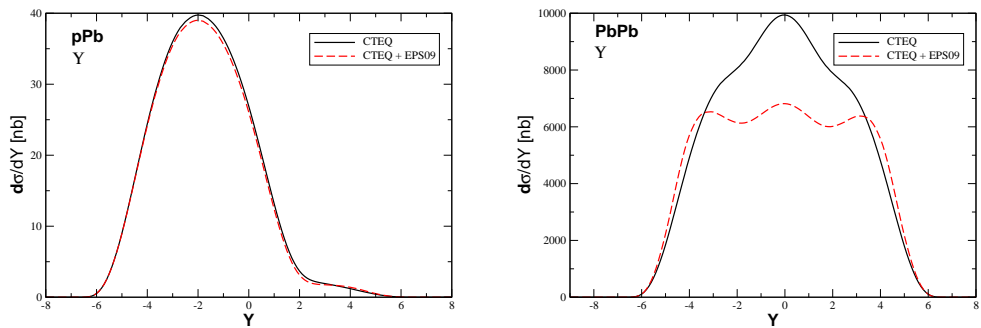

\includegraphics[scale=0.25]{pPb_ups.eps} 
\hspace{0.6cm}
\includegraphics[scale=0.25]{PbPb_ups.eps}
 \caption{(Color online). Rapidity distribution for the inelastic $\Upsilon$ photoproduction in coherent $pPb$ (left panel) and $PbPb$ (right panel) collisions at $\sqrt{s} = $ 5.5 TeV.  }
\label{fig7}
\vspace{1cm}
\end{figure*}

\begin{table}
\begin{center}
\begin{tabular} {||c|c|c||}
\hline
\hline
 & {\bf CTEQ} & {\bf CTEQ + EPS09}   \\
\hline
\hline
{\bf $J/\Psi$} & 16.1 $\mu$b ($2415 \times 10^3$ / $1610 \times 10^6$)   &  15.7	 $\mu$b ($2355 \times 10^3$ / $1570 \times 10^6$)    \\
\hline
{\bf $\Upsilon$} & 0.19 $\mu$b ($28 \times 10^3$ / $19 \times 10^6$)   &  0.18 $\mu$b ($27  \times 10^3$ / $18 \times 10^6$)   \\
\hline
\hline
\end{tabular}
\end{center}
\caption{The total cross section (event rates)  for the  inelastic quarkonium photoproduction in coherent $pPb$ collisions at  $\sqrt{s} = 5.5$ TeV.}
\label{tab3}
\end{table}

Lets discuss now the inelastic $J/\Psi$ photoproduction  in coherent $pPb$ collisions, considering that $h_1 = p$ and $h_2 = Pb$.   As discussed before, in this case we expect asymmetric rapidity distributions, with the contribution of the $\gamma p$ and $\gamma Pb$ interactions being different. In $\gamma p$ interactions the photon comes from the nuclei, with the photon flux being proportional to $Z^2$, and the photoproduction cross section being determined by the gluon distribution of the proton ($xg_p$). In $\gamma Pb$ interactions the photon comes from the proton and the photoproduction cross section being determined by the gluon distribution of the nuclei, which is enhanced by a factor of the order of $A = 208$ in comparison to $xg_p$.  In Fig. \ref{fig6} (left panel) we explicitly show the different contributions for the rapidity distribution for $\sqrt{s} = 5.5$ TeV  disregarding nuclear shadowing effects.  As expected, the $\gamma p$ contribution peaks for negative rapidities and $\gamma Pb$ one for positive rapidities, with the rapidity distribution being  asymmetric. In Fig. \ref{fig6} (right panel) we present our  predictions considering the nuclear shadowing effects, which are small, since the rapidity distributions is dominated by $\gamma p$ interactions. In Table \ref{tab3} we present our estimates for the total cross sections and production rates
assuming  the  design luminosity ${\cal L}^{\mathrm{pPb}}_{\mathrm{LHC}} = 150$ mb$^{-1}$s$^{-1}$ and a run time of $10^6$ seconds. 
We predict cross sections that are two orders of magnitude  smaller than those obtained in the $PbPb$ case. The larger $pA$ luminosity, which is two order of magnitude higher than for $AA$, counteracts this suppression for the event rates. However, the resulting events rates still are small in comparison to the $pp$ results. Recently, an upgraded $pPb$ scenario was proposed in Ref. \cite{david}, which improve the  $pPb$ luminosity and the running time. These authors proposed the following scenario for $pPb$ collisions:   ${\cal L}^{\mathrm{pPb}} = 10^4$ mb$^{-1}$s$^{-1}$  and a run time of $10^7$ s. The corresponding event rates also are  presented in the   Table \ref{tab3}. In this case we have  numbers similar to those for $pp$ collisions, which makes the experimental analysis feasible. Another advantage of $pPb$ collisions is that it is expected to trigger on and carry out the measurement with almost no pileup \cite{david}. Therefore, the upgraded $pA$ scenario provides one of the best possibilities to detect the inelastic $J/\Psi$ photoproduction in coherent processes.

Our predictions for the inelastic $\Upsilon$ photoproduction in coherent $pPb$ and $PbPb$ collisions are presented in Fig. \ref{fig7}. In comparison to the $J/\Psi$ case, the predictions are reduced by two orders of magnitude, which is directly associated to the larger mass of the $\Upsilon$. Moreover, the rapidity distribution is also reduced at midrapidity by the nuclear shadowing effects. The corresponding values for the cross sections and event rates are presented in the Tables \ref{tab2} and \ref{tab3}.

Finally, lets compare our predictions with those obtained  in  Ref. \cite{vicmag_mesons2} for the exclusive quarkonium production. Although in \cite{vicmag_mesons2} the numbers were obtained considering the color dipole formalism and taking into account saturation effects, similar predictions has been obtained using other approaches \cite{gluon,strikman,Schafer,gluon2,Schafer2}. Our predictions for the inclusive quarkonium photoproduction in $pp/pPb/PbPb$ collisions are a factor $\gtrsim$ 4 smaller than the exclusive $J/\Psi$ production. In the $\Upsilon$ case, the inclusive production is a factor $\gtrsim$ 3 smaller than the exclusive one.  It is important to emphasize that distinctly from the $J/\Psi$ case, which have its parameters constrained by the $ep$ HERA data, the predictions for the $\Upsilon$ production were obtained considering an educated guess for the parameters. This may be an explanation for the different values for the reduction factors  for $J/\Psi$ and $\Upsilon$ production.
Our results demonstrate that the contribution of the inelastic channel for quarkonium photoproduction is non-negligible. Moreover, if  the experimental separation of the inelastic quarkonium photoproduction in coherent interactions were possible, the large numbers obtained in our calculations indicate that this process could be used to study  
the underlying mechanism governing heavy quarkonium production.

\section{Conclusions}
\label{conc}

In this paper we have computed for the first time the cross sections for the inelastic quarkonium photoproduction  in coherent $pp/pPb/PbPb$ collisions at LHC energies. We predict sizeable values for the cross sections and event rates. In comparison with the exclusive production, the inelastic channel is non-negligible, being smaller by a factor $\gtrsim$ 3.
The experimental separation between the exclusive and inclusive quarkonium photoproduction still is an open question, in particular considering the large pileup event expected to occur at LHC. However, if the separation were possible, using for example forward detectors for the tagging of the hadrons in the final state, the inelastic quarkonium photoproduction in coherent interactions could be used to improve our understanding of the  mechanism of quarkonium production. 
Certainly our analysis deserves more detailed studies in several aspects, as for example the inclusion of next-to-leading order corrections to the Color Singlet Model, the study of differential distributions and the calculation of the quarkonium photoproduction using the NRQCD formalism. We plan to do these  studies in forthcoming publications.

\section*{Acknowledgments}
This work was supported by CNPq, CAPES and FAPERGS, Brazil.

\end{document}